\documentstyle[12pt,axodraw]{article}
\begin{document}
\pagestyle{empty} \vskip -2cm
\centerline{ hep-ph/0109151   \hfill  IMSc/2001/09/50}
%
\vskip 1cm
\begin{center}{\bf \large
Squark Mixing Contributions to CP violating phase $\gamma$ 
}

\vskip 1cm

{\large
Darwin Chang$^{a,b}$, We-Fu Chang$^{a,c}$, Wai-Yee Keung$^{d}$\\
Nita Sinha$^{e}$  and Rahul Sinha$^{e}$
}\\
{\em $^a$NCTS and Physics Department, National Tsing-Hua University,\\
Hsinchu 30043, Taiwan, R.O.C.}\\

{\em $^b$ Lawrence Berkeley Laboratory, 
         University of California, Berkeley, CA , USA}\\

{\em $^c$ TRIUMF, Vancouver, BC V6T2A3, Canada}\\

{\em $^d$Physics Department, University of Illinois at Chicago, 
         Illinois 60607-7059, USA}\\

{\em $^e$The Institute of Mathematical Sciences, 
         Taramani, Chennai 600 113, India}\\
\end{center}
\vskip .5cm
\begin{abstract}
We investigate the possibility that the CP violation due to the
soft supersymmetry breaking terms in squark mixing 
can give significant contributions to the various $\gamma$ related
parameters in $B$ decays, different from those of the 
Standard Model.  We derive the new
limits on $(\delta^u_{12})_{LL,LR,RR}$ and on
$(\delta^d_{23})_{LL,LR,RR}$ from the recent data on
$D^0$--$\bar{D}^0$  oscillation as well as those on  
$B_s^0$--$\bar{B_s}^0$ oscillation.  We show
that, together with all the other constraints, the currents limits
on these parameters still allow large contributions to the 
CP violating phases in  $B_s^0$--$\bar{B_s}^0$ as well as 
$D^0$--$\bar{D}^0$ oscillations which will modify some of the proposed
measurements of $\gamma$ parameters in CP violating $B$ decays.
However, the current constraints already dictate that 
the one-loop squark mixing contributions to various $B$ decay 
amplitudes cannot be competitive with that 
of the Standard Model (SM), at least for those $B$ decay modes which 
are dominated the tree level amplitudes within the SM, 
and therefore they are not significant in contributing to
CP asymmetries in the corresponding  $B$ decays.

\end{abstract}
\vskip 0.5cm
\centerline{PACS numbers:  14.80.Cp,  11.30.Er, 12.60.Jv, 14.40.Nd}
\newpage
%

\pagestyle{plain}
%

\section{Introduction}

With the $B$ factories producing physics at full steam, it is important
to investigate critically the possibility of distinguishing standard
Kobayashi-Maskawa (KM) \cite{km} model from the other alternatives of
CP violation.  In particular, one needs to investigate if there is any
non-KM mechanism that contributes to the measurement of CP violating
phases which within the KM context, are labeled as $\alpha$, $\beta$
and $\gamma$ (or $\phi_{1,2,3}$ in another popular convention in the
literature).  In fact, it is interesting to note that some of the CP
violating asymmetries in $B$ decays, which within the KM mechanism are
considered identical, may correspond to different numerical
asymmetries in an alternative theory of CP violation.

One of the leading extensions of the Standard Model is the
supersymmetric version of the theory.  The addition of supersymmetric
partners, as well as the necessity of supersymmetry breaking creates
large variations of such models.  In this paper we shall assume that
the spectrum of the supersymmetric extension is minimal in the sense
that no additional supermultiplet is introduced beyond the usual two
copies of the Higgs doublet needed for fermion masses.  In this
theory, the new parameters are the coupling constants related to soft
supersymmetry breakings.  It is well known that these soft breaking
parameters can give rise to new sources of CP violation.

Among these soft breaking parameters, the ones most relevant 
to CP violation in
$K$ or $B$ decays are the dim-2 soft squark mixing parameters
(the matrices $M^2_{\tilde{Q}}$ for left-handed squarks, and, $M^2_{\tilde{U}}$,
$M^2_{\tilde{D}}$ for right-handed squarks), as well as
the dim-3 trilinear scalar Yukawa couplings ($Y^A_u, Y^A_d$) which, after
the spontaneous symmetry breaking, creates the mixings between left- and
right-squarks (the matrices 
$Y^A_u \langle H_u\rangle $ and $Y^A_d \langle H_d \rangle$, here we follow
the notation of Ref.\cite{mm}).
There are many discussions in the literature about whether the new
supersymmetric contributions can give large enough $\epsilon$ and
$\epsilon'$ \cite{mm,epsilon,kn}.

In Ref.\cite{mm}, it was pointed out that even though the natural
value of $(\delta^d_{12})_{LR}$ in generic models may be of order
$10^{-5}$, its contribution is big enough to saturate the experimental
value for $\epsilon'_K$.  One may wonder whether it is possible to use
$(\delta^d_{12})_{LR}$ to saturate both $\epsilon_K$ and $\epsilon'_K$
in kaon system\cite{mm,brhlik2}.  However it was pointed out
\cite{Masiero} that in order to saturate both, the absolute value of
$(\delta^d_{12})_{LR}$ has to be about $3 \times 10^{-3}$ which is
larger than its generic value.

On the other hand, in Ref.\cite{kn}, it was pointed out that if one
takes into account the isospin breaking effect of the supersymmetric
$\Delta S= 1$ box diagrams, it is possible to account for $\epsilon'$
even using the $\hbox{Im}(\delta^d_{12})_{LL}$ with a mild fine-tuning.  Note
that, as emphasized in Ref.\cite{Masiero}, while $\delta_{LR}$'s
should be generically small due to its $SU(2)$ breaking character,
$\delta_{LL}$ and $\delta_{RR}$ do not have to be small.  This opens up
the possibility to use $(\delta^d_{12})_{LL}$ alone to saturate both
$\epsilon$ and $\epsilon'$.  

In Ref.\cite{masiero2}, it was pointed out that while it is possible
for $(\delta^d_{13})_{LR}$ to contribute to $\beta$ parameter in $B$
decays due to the $B^0_d$--$\bar B^0_d$ oscillation, 
however its generic value is 
typically too  small to account for the recent data from $B$ factories,
$\sin2\beta = 0.59\pm0.14\pm0.05$ (for Babar\cite{ref:babar}) 
and 
$\sin2\beta = 0.99\pm0.14\pm0.06$ (for Belle\cite{ref:belle}).

The purpose of this paper is to address the issue  whether these new
supersymmetric sources of CP violation can give rise to asymmetries which
are usually associated with $\gamma$.  We will consider only those 
$B$ or $B_s$ decay modes  whose amplitudes are dominated by tree 
amplitudes within Standard Model
(that is, the modes without ``Penguin pollution'').  We call
these decays, non-Penguin type.  We shall briefly comment on modes
with Penguin contributions later on.

We first derive constraints on squark mixing parameters based on the
most recent data on $B_s$--$\bar{B}_s$ and $D^0$--$\bar{D}^0$
oscillations.  Within SUSY, $\Delta B =1$ box diagrams also
arise. These can contribute even to modes that have only tree level
contributions within the SM. Using the constraints obtained on the
squark mixing parameters, we show that the contributions of $\Delta B
=1$ SUSY box diagrams to the these (non-Penguin) $B$ decays are
negligible.  However, we find that SUSY contribution can give rise to
CP asymmetries through either the initial state $B^0_s$--$\bar B^0_s$
oscillation or the final state $D^0$--$\bar D^0$ oscillation.  We
compare these asymmetries with the predictions of KM model.

\section{New Limits on Squark Mixings}

In a comprehensive paper\cite{Gabbiani}, Gabbiani et. al. work out
various limits on the flavor changing couplings
$(\delta^q_{ij})_{LL,LR,RR}$, where $q$ can be $u$(up-type),
$d$(down-type) or $\ell$(lepton), and $ij$ are generation indices.
The $\delta$'s are dimensionless parameters defined as
\begin{equation}
(\delta_{ij})_{AB} = { (m^2_{ij})_{AB}  /  m^2_{\tilde q}},
\end{equation}
where $AB$ and $ij$ stand for the chirality and flavor
respectively. For our purpose we will need only
$(\delta^u_{12})_{LL,LR,RR}$ and $(\delta^d_{23})_{LL,LR,RR}$. The
earlier limits on these parameters, given in Ref.\cite{Gabbiani}, are
summarized in Tables 1 and 2.

Recently there are new measurements on the $D^0$--$\bar{D}^0$
oscillation as well as on the $B_s^0$--$\bar{B_s}^0$ oscillation.  
They can be translated into new limits on these $\delta$ parameters.  
In particular, for $D^0$--$\bar{D}^0$ oscillation, 
the new measurements give $\triangle m_D< 0.461\times 10^{-10}$MeV 
(in Ref.\cite{PDG2000}). 
For $B_s$-$\bar B_s$, the data have so far not been able to set a 
solid upper limit on the oscillation frequency, $\Delta M_s$, due 
to the error in the measurement of the higher frequency region
\cite{boix}.  The combined data from LEP and SLD give 
$\Delta M_{B_s} > 15$ ps$^{-1}$ at 95 \% C.L. 
However, a hint of oscillation is observed (with large error)
around $\Delta m_{B_s}$ of 17 ps$^{-1}$ \cite{boix}. 
The new measurements from LEP and CDF can also be combined to give  
${\Delta \Gamma_s / \Gamma_s} = 0.16^{+0.08}_{-0.09}$, or
${\Delta \Gamma_s / \Gamma_s} < 0.31$, at  $95\%$ C.L. 
This can be combined with the lattice calculation of 
${\Delta \Gamma_s / \Delta M_{B_s}} = 3.5^{+0.94}_{-1.55} \times 
10^{-3}$\cite{hashimoto}  in SM, to give 
$\triangle m_{B_s}=(29^{+16}_{-21})  \hbox{ ps}^{-1}$ .
In presence of SUSY contributions, we expect, this lattice estimate to
change. To obtain reasonable limits on the $\delta_{23}$ parameter, 
we use the suggestive values of $\Delta m_{B_s}=8, 17, 45$ ps$^{-1}$ 
as typical value in our study.  Note that these are not yet serious 
experimental limits, however as commented later, our physical 
conclusions on $B$ decays in the next section, are not  significantly 
altered even if $\Delta M_{B_s}$ turns out to be one order of magnitude 
larger. 

$$  \begin{array}{||c|c|c|c||}  \hline \hline
                     & & &    \\
  x & \sqrt{\left|\Re  \left(\delta^{u}_{12} \right)_{LL}^{2}\right|}  &
 \sqrt{\left|\Re  \left(\delta^{u}_{12} \right)_{LR}^{2}\right|}  &
 \sqrt{\left|\Re  \left(\delta^{u}_{12} \right)_{LL}\left(\delta^{u}_{12}
 \right)_{RR}\right|} \\
                       & & & \\
 \hline
    0.3 & 4.7\times 10^{-2}  & 6.3\times 10^{-2}  & 1.6\times 10^{-2} \\
    1.0 & 1.0\times 10^{-1}  & 3.1\times 10^{-2}  & 1.7\times 10^{-2} \\
    4.0 & 2.4\times 10^{-1}  & 3.5\times 10^{-2}  & 2.5\times 10^{-2} \\  
 \hline \hline
 \end{array} $$
Table 1: Limits on $\mbox{Re}\left(\delta_{12}^u\right)_{AB}
 \left(\delta_{12}^u\right)_{CD}$ from $\triangle m_D$, 
with $A,B,C,D=(L,R)$, for an average squark mass
 $m_{\tilde{q}}=500\mbox{GeV}$ and for different values of
$x=m_{\tilde{g}}^2/m_{\tilde{q}}^2$.
$$  \begin{array}{||c|c|c||}  \hline \hline
  & & \\
 x & \left|\left(\delta^{d}_{23} \right)_{LL}\right|  &
     \left|  \left(\delta^{d}_{23} \right)_{LR}\right|  \\
  & & \\ \hline
   0.3  & 4.4  & 1.3\times 10^{-2}   \\
   1.0  & 8.2  & 1.6\times 10^{-2}   \\
   4.0  & 26   & 3.0\times 10^{-2}  \\ \hline \hline
 \end{array} $$
Table 2: Limits on the $\left| \delta_{23}^{d}\right|$ from
$b\rightarrow s \gamma$ decay for an average squark mass 
$m_{\tilde{q}}=500\ \mbox{GeV}$ and for different values of 
$x=m_{\tilde{g}}^2/m_{\tilde{q}}^2$. For different values of $m_{\tilde{q}}$,
the limits can be obtained multiplying the ones in the table by
$(m_{\tilde{q}}/500\ \mbox{ GeV})^2$.

For  $B^0_s$--$\bar B_s^0$ oscillation, the $(\delta^d_{23})_{ab}$ mixing
contributes
to the operator
\begin{eqnarray}
    Q_1 & = & 
     \bar s^{\alpha}_L  \gamma_\mu b^{\alpha}_L 
     \bar s^{\beta}_{L} \gamma^\mu b^{\beta}_L\; ,
    \nonumber \\
    Q_2 & = & \bar s^{\alpha}_R  b^{\alpha}_L \bar s^{\beta}_R b^{\beta}_L\; ,
    \nonumber \\
    Q_3 & = & \bar s^{\alpha}_R  b^{\beta}_L \bar s^{\beta}_R b^{\alpha}_L\; ,
    \nonumber \\
    Q_4 & = & \bar s^{\alpha}_R  b^{\alpha}_L \bar s^{\beta}_L b^{\beta}_R\; ,
    \nonumber \\
    Q_5 & = & \bar s^{\alpha}_R  b^{\beta}_L \bar s^{\beta}_L b^{\alpha}_R\; ,
    \label{Qi}
\end{eqnarray}
plus the operators $\tilde{Q}_{1,2,3}$ obtained from the $Q_{1,2,3}$ by the
exchange $ L \leftrightarrow R$.
Here $q_{R,L}={1\over2}(1\pm\gamma_5) q$, and $\alpha$ and $\beta$ are color
indices. The color matrices normalization is 
$\mathop{\mbox{Tr}} (t^A t^B)={1\over2} \delta^{AB}$.
The $\Delta B=2$ effective Hamiltonian reads:
\begin{eqnarray}
    \cal{H}_{\mbox{eff}}&=&-\frac{\alpha_s^2}{216 m_{\tilde{q}}^2}\Biggl\{
    \left(\delta^d_{23}\right)^2_{LL}
    \left(  24\,Q_1\,x\,f_6(x) + 66\,Q_1\,\tilde{f}_6(x) \right)
    \nonumber \\
    &+&  \left(\delta^d_{23}\right)^2_{RR}
    \left(  24\,\tilde{Q}_1\,x\,f_6(x) + 66\,\tilde{Q}_1\,
    \tilde{f}_6(x) \right)
    \nonumber \\
    &+&  \left(\delta^d_{23}\right)_{LL}\left(\delta^d_{23}\right)_{RR}
    \left( 504\,Q_4\,x\,f_6(x) - 72\,Q_4\,\tilde{f}_6(x)
    \right. \nonumber \\
    &&+ \left. 24\,Q_5\,x\,f_6(x) + 120\,Q_5\,\tilde{f}_6(x) \right)
    \nonumber \\
    &+&  \left(\delta^d_{23}\right)^2_{RL}
    \left(  204\,Q_2\,x\,f_6(x) - 36\,Q_3\,x\,f_6(x) \right)
    \nonumber \\
    &+&  \left(\delta^d_{23}\right)^2_{LR}
    \left(  204\,\tilde{Q}_2\,x\,f_6(x) - 36\,\tilde{Q}_3\,x\,f_6(x) \right)
    \nonumber \\
    &+&  \left(\delta^d_{23}\right)_{LR}\left(\delta^d_{23}\right)_{RL}
    \left( - 132\,Q_4\,\tilde{f}_6(x) - 180\,Q_5\,\tilde{f}_6(x)
    \right)  \Biggr\},
    \label{eq:box}
\end{eqnarray}
where $x=m^2_{\tilde{g}}/m_{\tilde{q}}^2$, $m_{\tilde{q}}$
is the average squark mass involved in the box diagram,
$m_{\tilde{g}}$ is the gluino mass and the functions $f_6(x)$ and
$\tilde{f}_6(x)$ are given by :
\begin{eqnarray}
f_6(x)=\frac{6(1+3x)\ln x +x^3-9x^2-9x+17}{6(x-1)^5}\; , \nonumber  \\
\tilde{f}_6(x)=\frac{6x(1+x)\ln x -x^3-9x^2+9x+1}{3(x-1)^5}\; .
\end{eqnarray}
Note that $f_6(x=1)=1/20$ while $\tilde{f}_6(x=1)=-1/30$, 
therefore they cancel a lot in the combination 
$24 x f_6(x) + 66 \tilde{f}_6(x) = -1$ for $x=1$.  

The matrix elements of the operators $Q_1$, $Q_2$  are defined as
\begin{eqnarray}\label{qb}
\langle\bar B_s|Q_1|B_s\rangle &=& \frac{2}{3}f^2_{B_s}M^2_{B_s} 
{\cal B}, \\
\label{qsbs}
\langle\bar B_s|Q_2|B_s\rangle &=& -\frac{5}{12}f^2_{B_s}M^2_{B_s}
\frac{M^2_{B_s}}{(\bar m_b+\bar m_s)^2} {\cal B}_S, \nonumber\\
 &\equiv& -\frac{5}{12}f^2_{B_s}M^2_{B_s}{\bar{\cal B}_S}~.   
\end{eqnarray}
The matrix elements of the other operators in Eq.(\ref{Qi}) can be
obtained in terms of those given in Eqs.(\ref{qb},\ref{qsbs}).  The
bag factors ${\cal B}$ and ${\cal B}_S$ parameterize the non
perturbative contributions to the matrix elements and have been
evaluated \cite{beneke} on the lattice to be ${\cal B} =0.9\pm 0.1$
and ${\bar{\cal B}_S}= 1.25\pm 0.1$.

%

Similar equations apply to the calculation of the mass difference of
$D^0$-$\bar D^0$ system.  
Using the same naive estimate of hadronic matrix element
used in Ref.\cite{Gabbiani}, the results may be  summarized in Table 3.
Note that, while for the $B_s$, the hadronic matrix elements discussed
above, have been obtained with some rigor, those for the $D$ meson
assume an universal bag parameter of unity.

As one can see by comparing Table 3 with Table 1 and Table 2, 
the recent (and coming) data do provide significant improvement on the limits 
on $\left(\delta^u_{12}\right)_{AB}$.

$$ \begin{array}{||c|c|c|c||}  \hline \hline
 & & &  \\
 x & \sqrt{\left|\Re  \left(\delta^{u}_{12} \right)_{LL}^{2}\right|} 
   & \sqrt{\left|\Re  \left(\delta^{u}_{12} \right)_{LR}^{2}\right|}  &
 \sqrt{\left|\Re  \left(\delta^{u}_{12} \right)_{LL}\left(\delta^{u}_{12}
 \right)_{RR}\right|}  \\
 & & &  \\
 \hline
    0.3  & 2.58\times 10^{-2}  & 3.43\times 10^{-2}  & 8.52\times 10^{-3}  \\
    1.0  & 5.46\times 10^{-2}  & 1.72\times 10^{-2}  & 9.49\times 10^{-3}  \\
    4.0  & 1.28\times 10^{[B-1}  & 1.90\times 10^{-2}  & 1.34\times 10^{-2}  \\
  \hline \hline 
 & & & \\
 x & \sqrt{\left|\Re  \left(\delta^{d}_{23} \right)_{LL}^{2}\right|} 
   & \sqrt{\left|\Re  \left(\delta^{d}_{23} \right)_{LR}^{2}\right|}  &
 \sqrt{\left|\Re  \left(\delta^{d}_{23} \right)_{LL}\left(\delta^{d}_{23}
 \right)_{RR}\right|}  \\
 & & & \\
 \hline
    0.3  & 0.16, 0.23, 0.38  & 0.20, 0.29, 0.46  & 0.06, 0.08, 0.13  \\
    1.0  & 0.34, 0.50, 0.81  & 0.11, 0.17, 0.27  & 0.06, 0.09, 0.15  \\ 
    4.0  & 0.80, 1.17, 1.90  & 0.13, 0.18, 0.30  & 0.09, 0.13, 0.21  \\ 
 \hline \hline
 \end{array}   $$
Table 3: Values of  
$\mbox{Re}\left(\delta_{ij}\right)_{AB} \left(\delta_{ij}\right)_{CD}$,
with $A,B,C,D=(L,R)$.
The upper part of the table are derived 
from saturating $\triangle m_D< 0.479\times 10^{-10}$ MeV
by squark mixing contribution. 
The lower part of the table are based on suggestive values 
$\triangle m_{B_s}=8, 17, 45  \hbox{ ps}^{-1}$. 
We use  an average squark mass $m_{\tilde{q}}=500$ GeV and 
choose different values of
$ x=m_{\tilde{g}}^2/m_{\tilde{q}}^2$.  The constraints on  
$\left(\delta_{ij}\right)_{RR}$ are the same as those on 
$\left(\delta_{ij}\right)_{LL}$ in the Table.

\section{SUSY contributions to $\gamma$}

The clean measurement of $\gamma$ has been a challenge, leading to
several attempts at providing feasible techniques to measure it.  SUSY
contributions to $\gamma$ therefore vary, depending on the method used.
We therefore first discuss the various methods proposed to measure $\gamma$.
The original suggestion \cite{GLW} for cleanly measuring $\gamma$
involved the decays 
$B^{\pm} \rightarrow \stackrel{ {\tiny (} \hbox{\small ---} {\tiny )} } {D^0}K^{\pm}$
and $D^0_{CP} K^{\pm}$, where $D^0_{CP}$ stands for the CP eigenstate of $D$.
However, since it is virtually impossible to tag the flavor of the
$D$ meson, the method was improved.
Ref.~\cite{ADS} considered the 
$\stackrel{ {\tiny (} \hbox{\small ---} {\tiny )} } {D^0}$ produced, 
to subsequently decay
to at least two final states. 
The mode 
$B^{\pm} \rightarrow 
{\stackrel{ {\tiny (} \hbox{\small ---} {\tiny )} } {D^{0*}}} {K^*}^{\pm}$ 
was proposed in Ref.~\cite{SS}. For the purpose of this
paper the arguments made to the generic mode $B^{\pm} \rightarrow
\stackrel{ {\tiny (} \hbox{\small ---} {\tiny )} } {D^0} K^{\pm}$ 
applies to all the methods in Ref.\cite{GLW,ADS,SS}.
Alternative modes involving $B_s$ mesons have also been suggested to
measure $\gamma$. They include $B_s^0/\overline{B_s^0}\rightarrow
D_s^\mp K^\pm$ \cite{ADK} (see Fig.~1) 
and its final state vector meson analogue
$B_s^0/\overline{B_s^0}\rightarrow {D_s^*}^\mp K^{*\pm}$\cite{LSS}.
One may note that all the modes discussed above only have tree level
contributions.  Other methods involving $B\to K\pi$ together with
$B\to KK$, etc., which include penguin contribution, have also been
considered\cite{Gronau-review}. However, they involve theoretical
assumptions like the inherent use of SU(3) or factorization
assumption.

All CP asymmetries arise from a relative phase between two decay
channels to the same final state. This may arise either due to two or
more contributions to the direct decay or, due to the oscillation of the
initial or final state neutral meson.

Let us consider the CP asymmetry appearing in charged decay modes
$B^{\pm} \rightarrow D^0 K^{\pm}$ or $\bar{D}^0 K^{\pm}$.  In SM, the
first stage of the decay subprocess at quark level is either due to $b
\rightarrow u(\bar{c}s)$, which is severely Cabbibo suppressed but
complex in the Wolfenstein (or Chau-Keung) convention\cite{wck}, or,
due to $b \rightarrow c (\bar{u}s)$ which is doubly Cabbibo suppressed
and real in the same convention.  By observing decays of $D^0$ and
$\overline{D}^0$ to a common final state $f$ (which may be a CP
eigenstate), one achieves the required interference. The relevant
effective Hamiltonian for the quark level process $b \rightarrow
u(\bar{c}s)$, in SM is
\begin{equation}
\hbox{$1\over2$} (g_2^2/M_W^2) V_{cs}^*V_{ub}
     (\bar u_L \gamma_\mu  b_L) (\bar s_L \gamma^\mu  c_L) \ ,
\end{equation}
where $V_{cs}^*V_{ub} \sim A \lambda^3 e^{-i \gamma}$; while for the quark level process $b \rightarrow c (\bar{u}s)$, it is, 
\begin{equation}
\hbox{$1\over2$} (g_2^2/M_W^2) V_{us}^*V_{cb}
     (\bar c_L \gamma_\mu  b_L) (\bar s_L \gamma^\mu  u_L) \ ,
\end{equation}
where $V_{us}^*V_{cb} \sim A \lambda^3$.
The relative phase gives rise to CP violating parameter 
$\gamma$ directly.

In the scenario of squark mixing, the effective Hamiltonian for 
$b\to u (\bar c s)$  is
\begin{equation}
   {\cal H}_{\rm eff}^{b\to u(\bar c s)}
  = - \frac{\alpha_s^2}{108 m_{\tilde{q}}^2}
    (\delta^u_{12})_{LL} (\delta^d_{23})_{LL}
    \left(  24      x\,f_6(x) 
          + 66 \tilde{f}_6(x) 
    \right)
     (\bar u_L \gamma_\mu  b_L) (\bar s_L \gamma^\mu  c_L) + \cdots
  \ ,
\end{equation}
with $x=m^2_{\tilde{g}}/m_{\tilde{q}}^2$.
We only list the contribution from the channel $LL$ in chirality. 
Other amplitudes due to the insertion of other $\delta$ parameters 
are easily obtained from Eq.~(\ref{eq:box}).

For SUSY contribution to the $\Delta B = 1$ amplitude to be relevant 
to CP asymmetry, it has to be a sizable contribution to the decay 
amplitude.  To estimate the SUSY contribution we take $x=1$,  
$24 x\,f_6(x)   + 66 \tilde{f}_6(x) = -1 $ and 
optimistically use $\alpha_s^2 \sim 0.02$
$|V_{ub}| \sim 0.003$, $ (\delta^u_{12})_{LL} \sim 0.06$,
$(\delta^d_{23})_{LL} \sim 0.7$.
The amplitude ratio of the SUSY to the SM is about
\begin{equation}
{ (\alpha^2_s/108) 
 10^{-4}\hbox{ GeV}^{-2}(100\hbox{ GeV}/M_{\tilde{q}})^2 
\times 0.06 \times 0.7
     \over 
 (4 G_F/\sqrt{2})\times  0.003  } 
\sim    \left({100\hbox{ GeV} \over M_{\tilde{q}} }\right)^2 \times 10^{-2}
\ . \label{eq:sqconstraint} \end{equation}
Assuming that the SUSY $\Delta B=2$ box diagrams dominate in the
$B_s-\overline{B}_s^0$ and $D^0-\overline{D}^0$ oscillations, the same
ratio can be expressed more directly in terms of measured quantities as,
\begin{equation}
  {3 \sqrt{\Delta M_D \Delta M_{B_s} } \over 
  f_D f_{B_s} \sqrt{       M_D        M_{B_s} } }
    \left/
  { 4 G_F |V_{ub}| \over \sqrt{2}} \right. \sim 10^{-4} 
\  .  \label{eq:sqnumber} \end{equation}
The numerical result is based on the central values of  the parameters,
$\Delta M_D    = 0.07$ ps$^{-1}$,
$ f_D=0.2$ GeV, and
$ f_{B_s}=0.23$ GeV, as well as 
the suggestive value $\Delta M_{B_s}=45 $ ps$^{-1}$.
If one takes $M_{\tilde q}=500$ GeV in Eq.~(\ref{eq:sqconstraint}), 
The number comes up to be $10^{-4}$ as in Eq.~(\ref{eq:sqnumber}).
Therefore even after taking the parameters to be favorable to the 
SUSY contribution, such one loop contribution is still much smaller 
than the highly KM suppressed SM tree amplitude.  
(Even if the color suppression of the contribution of the KM 
operator to $B^+ \rightarrow D^0 K^+$ is taken into account, SUSY 
is still very much a minor contribution to decay amplitudes).
Therefore we conclude that SUSY contribution to the $\Delta B=1$ 
box diagram (Fig.~2) is irrelevant to the CP asymmetry as long as we limit 
our consideration to $\delta^d_{23}$ and $\delta^u_{12}$.  
The same conclusion can be applied to all the $B$ decay modes.  

One may hence conclude that the only possible SUSY contributions to
the CP asymmetries arise from initial state $\Delta B=2$ or final
state $\Delta D=2$ transition, or both. Among the various methods to
determine $\gamma$, $B_s\to D_s^\mp K^\pm$ get contributions from
initial state $B_s^0-\overline{B}_s^0$ oscillation, 
$B^\pm\to \stackrel{ {\tiny (} \hbox{\small ---} {\tiny )} } {D^0} K^\pm$
get contributions from final state $D^0-\overline{D}^0$ oscillation and
and 
$B_s\to\stackrel{ {\tiny (} \hbox{\small ---} {\tiny )} } {D^0} \phi$ 
get contributions from both
$B_s^0-\overline{B}_s^0$ and $D^0-\overline{D}^0$ oscillations.

We first consider the contributions from initial state
$B_s^0-\overline{B}_s^0$ oscillation. 
Since SUSY particles are heavy, 
we assume that $\Gamma_{12}$ is not modified by SUSY and 
parameterize the SUSY contributions by  
$M_{12}^{SUSY}=M_{12}^{SM} y e^{i\eta}$. Hence, we have, 
\begin{equation}
  \label{eq:x-and-eta}
  \frac{\Gamma_{12}}{M_{12}}={    \Gamma_{12}^{SM} / M_{12}^{SM} \over
   1+M_{12}^{SUSY} /M_{12}^{SM} }=\frac{s e^{i\phi}}{1+y e^{i \eta}}  \ .
\end{equation}
$s$ and $\phi$ are SM parameters, with $s\sim O(10^{-2})$ and
$\phi\approx 0$. In terms of the SUSY parameters discussed earlier,
for the $LL$ chirality we have,
\[
y= \frac{\alpha_s^2}{216 m_{\tilde{q}}^2}
{\frac{1}{3} M_{B_s} f_{B_s}^2 {\cal B}_S \over M_{12}^{SM}}
    \left(  24\,x\,f_6(x) + 66\,\tilde{f}_6(x) \right)
 \bigg|\left(\delta^d_{12}\right)^2_{LL}\bigg|~.\]
The expression for $M_{12}^{SM}$ may be taken from Ref.\cite{buras}.
 Using,
\begin{equation}
  \label{eq:delta-mass}
  \Delta M= -2 {\rm Re} \Big(\frac{q}{p}(M_{12}
            -\frac{i}{2}\Gamma_{12})\Big)~,
\end{equation}
 the expressions for $\Delta M$  including
SUSY may be written as,
\begin{eqnarray}
  \label{eq:full-deltaM}
  \Delta M &=& 2 |M_{12}^{SM}|\;{\rm Re}\Bigg(1+2 y\cos\eta+y^2-\frac{s^2}{4}
-i s \Big(\cos\phi+ y  \cos(\phi-\eta)\Big)
  \Bigg)^\frac{1}{2}\nonumber\\
&=& 2 |M_{12}^{SM}|\;\sqrt{ (1+2
          y\cos\eta+y^2 )}\: (1+ O(s^2))  
\end{eqnarray}
The effect of SUSY on the meson mixing parameter $(q/p)$ can be expressed as,
\begin{equation}
  \label{eq:q-by-p}
   \Bigg(\frac{q}{p}\Bigg)^2=\frac{{M_{12}^{SM}}^*}{M_{12}^{SM}}
    \frac{(1+y e^{-i \eta})}{(1+y e^{i \eta})}  \ .
\end{equation}
Since, $M_{12}^{SM}$ is real for $B_s$, the argument $\theta_{B_s}$ of
$({q}/{p})$ is given by,
\begin{equation}
\label{theta}
  \theta_{B_s}=\frac{1}{2} \tan^{-1}\Bigg( \frac{-y^2 \sin(2 \eta)-2 y\sin\eta}
 {1+y^2 \cos(2 \eta)+2 y\cos\eta}\Bigg).
\end{equation}
$y$ may be solved using Eq.~(\ref{eq:full-deltaM}) as a function of
$\eta$, leading to the oscillation phase 
$\theta_{B_s}$ as a function of $\eta$.
Using the upper limit of $\Delta M_{B_s}=45\,{\rm ps}^{-1}$
and theoretical estimate of $M_{12}^{SM}$, from Ref.~\cite{buras}, 
we find that  values of $\theta_{B_s}$ are allowed in the full range 
$(-45^\circ, +45^\circ)$. 
If, however, $M_{12}^{SUSY}\gg M_{12}^{SM}$, i.e. in the limit, $y\to
\infty$, we get $\theta_{B_s}= -\eta=- \hbox{ Arg
  }\big((\delta^d_{23})^2_{AA}\big)$, with $A= L$ or $R$.

The mode $B_s\to D_s^\mp K^\pm$ has been proposed to measure
$\sin^2\gamma$ (see Fig.~1).
In the presence of SUSY contributions to
$B_s^0-\overline{B}_s^0$ oscillation, the angle $\gamma$ gets modified,
$\gamma\to\gamma^\prime=\gamma_{KM}\pm\theta_{B_s}$, where, $\theta_{B_s}$
is given by Eq.(\ref{theta}) and $\gamma_{KM}$ is the contribution from Kobayashi-Maskawa phase.  
The contribution from SUSY
to both the modes $B_s\to D_s^\mp K^\pm$ and $B_s\to
(\psi\!/\!J)\phi$ are identical.
This is very much different from the KM predictions
in which the asymmetry in $B^0_s \rightarrow \psi/J +\phi$ is
negligible while that is $B^0_s \rightarrow D_s^- K^+$ is large.
Note that the phase $\theta_{B_s}$ can be measured 
directly using 
the mode $B_s\to (\psi/J)\phi$, which has a large branching ratio and 
should be easier to measure.

In decays of the type $B^{\pm} \rightarrow D^0 K^{\pm}$, as discussed
earlier, $\gamma$ is measured by utilizing a possibility of interference
between the two quark level processes $b\to c \bar{u}s$ and $b\to u
\bar{c} s$. Along the decay chain the $c$ or $\bar{c}$ produce in the
final state a $D^0$ or $\overline{D}^0$ mesons respectively. The two
contributions are added and interfere if both $D^0$ or $\overline{D}^0$ 
decay to the same final state $f_D$ and have a relative phase $\gamma$.  
Here, $f_D$ is one of the states that both $D$ and
$\bar{D}$ can decay into, such as $K^- \pi^+$ or CP eigenstates $K^+
K^-$, $\pi^+ \pi^-$, $K_s \pi^0$ or $K_s \phi$.  In fact the whole
discussion can be applied to the modes in which final state $K^+$ is
replaced by $\pi^+$ or $\rho^+$.
In the presence of $D^0$--$\bar D^0$ oscillation there are additional
contributions to this process.  As discussed in details in
Ref.\cite{Bigi,silva}, there are many different types of CP violation
that can manifest themselves in such decays. 
SUSY can have
potentially large contribution to the $D^0$--$\bar D^0$ oscillation, even
providing a large phase to the oscillation.  One can estimate the
$D^0$--$\bar D^0$ oscillation phase, $\theta_D$, by repeating the procedure
used to determine $\theta_{B_s}$, except that here
$\Gamma_{12}^{SM}/M_{12}^{SM}$ cannot be ignored.  In the SM model,
$\gamma$ is large, and $\theta_D$ is small. However, if SUSY were to
dominate, $\theta_{D}=- \hbox{ Arg}\big((\delta^u_{12})_{AB}
(\delta^d_{12})_{CD}\big)$, with A,B,C,D=(L,R), could give large
contributions to CP asymmetries even if $\gamma$ is small,
especially $\hbox{ Arg}\big((\delta^u_{12})^2_{LL}$ or 
$\hbox{ Arg}\big((\delta^u_{12})^2_{RR}$.
Ref.\cite{silva} has considered the various possible CP violating
asymmetries in detail for arbitrary $\theta_D$.
First of all, there is the phase in the $B^+$ decay
amplitude, which in KM model, is exactly the $\gamma$ parameter.  Then
there is the phase in the $D^0$-$\bar{D}^0$ oscillation,
$\theta_D= \hbox{ arg}(q_D/p_D)$ (here, $q_D$ $p_D$ are
the composition amplitudes in the $D$-$\bar{D}$ system).  In
addition, the differences in strong final state phase shifts in $B^+$
decays ($\Delta_B$) and in D decays ($\Delta_D$) may become relevant
for some CP observables.
There is CP violation of the type proportional to
$\sin \gamma \sin\Delta_B$
which is due to the CP asymmetry in the decays
$B^\pm \rightarrow D K^\pm$,
the final state phase shift is needed to produce the CP asymmetry for
charged $B$ decay as expected.
There is CP violation proportional to
$\sin \theta_D \sin\Delta_D$
which is similar to  the CP asymmetry in the decays
$D \rightarrow f_D$.
There is CP violation proportional to
$\sin \theta_D \cos\Delta_D$
which is due to $D$-$\bar{D}$ oscillation.
Finally, there is CP violation of the type proportional to
$\sin(\gamma+\theta_D) \cos\Delta_B$
which is due to interference between $B \rightarrow D$ decays and the
subsequent $D$-$\bar{D}$ mixing.  Last category is of course most
interesting.  In the KM model, $\gamma$ is large, however $\theta_D$
is small, while in the SUSY model we consider, $\gamma$ is small but
$\theta_D= \hbox{arg}((\delta^u_{12})^2_{AA})$
can be quite large (here, $(AA)$ can be either $LL$ or $RR$).
Of course, in the fourth type of CP
violation listed above, $\theta_D$ in SUSY model can duplicate the
effect of $\gamma$ in KM model, however as discussed in
Ref.\cite{silva} in details, there are enough CP asymmetries that one
can measure to distinguish the two contributions in principle.

Finally, we have so far avoided discussing $B$ decays that can receive
significant contributions from penguin or chromo-dipole moment types
of diagrams such as, at quark level, $b \rightarrow d \bar{s} s$, $b
\rightarrow s \bar{s} s$, $b \rightarrow d \bar{d} d$ or $b
\rightarrow s \bar{d} d$, either in KM or in SUSY models.  They
contribute partially to $B \rightarrow \pi \pi$, $B \rightarrow K
\pi$, $B \rightarrow K \phi$ and other processes.  While the $\Delta
B=1$ box diagram SUSY contribution is still negligible for these
processes as long as one considers only $\delta^d_{23}$ and
$\delta^u_{12}$ parameters, as discuss in Ref.\cite{ciuchini}, the
SUSY penguin and/or chromo-dipole moment types of contributions can
play important role in their CP asymmetries.  This issue is discussed
in Ref.\cite{mm2}.

\section{Conclusion}

We list below the  results of our study of 
$(\delta^d_{LL, RR})_{23}$ and $(\delta^u_{LL, RR})_{12}$ on the CP 
asymmetries in $B_d$ or $B_s$ decays.  
\begin{itemize}
\item[(I)] 
We use the stringent $D^0$--$\bar D^0$ mixing data to obtain tighter
limits on $(\delta^u_{LL,RR})_{12}$.  Based on suggestive values of 
$B_s$--$\bar B_s$\cite{boix}, we illustrate the amount of improvement 
can be made on the constraint of  $(\delta^d_{LL,RR})_{23}$ in the 
near future. 
\item[(II)] 
For CP asymmetry in $B$ decays, 
SUSY can give large contribution to $B_s$ decays due to 
$B_s$--$\bar B_s$ oscillation,
but cannot give large contribution to the complex phase
in the decay amplitude via $\Delta B=1$ box diagram.  
Thus SUSY  produces  CP asymmetries  different 
from those in the KM model.  For example, while in KM 
the $B^\pm\to \stackrel{ {\tiny (} \hbox{\small ---} {\tiny )} } {D^0} K^\pm$
process has large phase $\gamma$  
and $B_s \rightarrow (\psi/J) \phi$  has 
negligible CP violating phase, however  in SUSY, 
both processes receive the same CP phase  
due to the $B_s$--$\bar B_s$ oscillation.
\item[(III)] For $B_d$ decay modes or charged $B^\pm$ decays modes,
the CP asymmetries due to the above $\delta$'s are negligible.
However an exception is 
$B^\pm\to \stackrel{ {\tiny (} \hbox{\small ---} {\tiny )} } {D^0} K^\pm$, 
where there can be
  SUSY contribution to the CP asymmetry due to the final state
  $D^0$--$\bar D^0$ oscillation.
\end{itemize}

\noindent {\bf Acknowledgments} This work was partially supported in
parts by U.S. Department of  Energy (grant number DE-FG02-84ER40173) 
and by National Science Council of R.O.C. 
The work of NS was supported by the  Young Scientist scheme of the
Department of Science and Technology, India.
DC wishes to thank H. Murayama, A. Masiero and L. Wolfenstein for 
discussions and SLAC Theory Group for hospitality.
NS and RS thank National Center for Theoretical Physics in 
Taiwan for hospitality during the visit when this work originated. 
\bibliographystyle{unsrt}

%
%
\section*{Standard Model Contribution}
\def\Xsize{220} \def\Ysize{50}
\def\Xoff{-30} \def\Yoff{60}

\begin{picture}(\Xsize,\Ysize)(\Xoff,\Yoff)
\GOval(50,50)(30,10)(0){0.5}
\ArrowLine(55,70)(95,70) \Text(75,78)[b]{$b$}
\ArrowLine(95,30)(55,30) \Text(75,25)[t]{$s$}
\Line(10,52)(40,52) \Text(15,55)[b]{$\bar B_s(b\bar s)\to$}
\Line(40,48)(10,48)
\ArrowLine(95,70)(135,50) \Text(115,55)[t]{$u$}
\ArrowLine(135,30)(95,30) \Text(115,25)[t]{$s$}
\Photon(95,70)(110,85){3}{3} \Text(100,80)[br]{\tiny  W}
\GOval(140,40)(15,7)(0){0.5} \Text(147,40)[l]{$\to K^+(\bar s u)$}
\ArrowLine(135,73)(110,85) \Text(120,75)[t]{$c$}
\ArrowLine(110,85)(135,97) \Text(120,95)[b]{$s$}
\GOval(140,85)(15,7)(0){0.5} \Text(147,85)[l]{$\to D_s^-(\bar cs)$}
\Text(95,0)[b]{$\sim V_{ub}V_{cs}^* \sim A\lambda^3e^{-i\gamma}$}
\end{picture}
\begin{picture}(\Xsize,\Ysize)(\Xoff,\Yoff)
\GOval(50,50)(30,10)(0){0.5}
\ArrowLine(95,70)(55,70) \Text(75,78)[b]{$b$}
\ArrowLine(55,30)(95,30) \Text(75,25)[t]{$s$}
\Line(10,52)(40,52) \Text(15,55)[b]{$B_s(\bar b s)\to$}
\Line(40,48)(10,48)
\ArrowLine(135,50)(95,70) \Text(115,55)[t]{$c$}
\ArrowLine(95,30)(135,30) \Text(115,25)[t]{$s$}
\Photon(95,70)(110,85){3}{3} \Text(100,80)[br]{\tiny  W}
\GOval(140,40)(15,7)(0){0.5} \Text(147,40)[l]{$\to D_s^-(\bar cs)$}
\ArrowLine(110,85)(135,73) \Text(120,75)[t]{$u$}
\ArrowLine(135,97)(110,85) \Text(120,95)[b]{$s$}
\GOval(140,85)(15,7)(0){0.5} \Text(147,85)[l]{$\to K^+(\bar s u)$}
\Text(95,0)[b]{$\sim V^*_{cb}V_{us} \sim A\lambda^3$}
\end{picture}

\vskip 4cm
\begin{itemize}
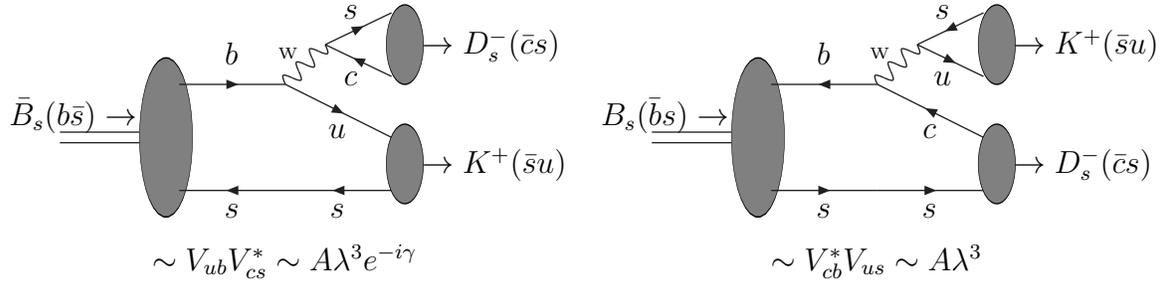
\item[Fig.~1] Quark diagrams for the tree-amplitudes in SM for the 
process $(B_s/\bar B_s)\to K^+D_s^-$.
\end{itemize}

\section*{Squark Mixing Contribution}

\begin{picture}(\Xsize,\Ysize)(\Xoff,\Yoff)
\GOval(50,50)(30,10)(0){0.5}
\ArrowLine(55,70)(95,70) \Text(75,78)[b]{$b$}
\ArrowLine(95,30)(55,30) \Text(75,25)[t]{$s$}
\Line(10,52)(40,52) \Text(15,55)[b]{$\bar B_s(b\bar s)\to$}
\Line(40,48)(10,48)
\ArrowLine(108,63)(135,50) \Text(115,55)[t]{$u$}
\ArrowLine(135,30)(95,30) \Text(115,25)[t]{$s$}

\DashLine(95,70)(110,85){2} \Text(102,78)[]{+}
\Photon(95,70)(108,63){2}{2.5} \Line(95,70)(108,63)
\Photon(110,85)(123,79){2}{2.5} \Line(123,79)(110,85)
\DashLine(108,63)(123,79){2}   \Text(116,71)[]{+}

\GOval(140,40)(15,7)(0){0.5} \Text(147,40)[l]{$\to K^+(\bar s u)$}

\ArrowLine(135,73)(123,79) \Text(130,70)[t]{$c$}
\ArrowLine(110,85)(135,97) \Text(120,95)[b]{$s$}
\GOval(140,85)(15,7)(0){0.5} \Text(147,85)[l]{$\to D_s^-(\bar cs)$}
\Text(95,0)[b]{$\sim \delta^d_{23}\delta_{12}^u$}
\end{picture}
\begin{picture}(\Xsize,\Ysize)(\Xoff,\Yoff)
\GOval(50,50)(30,10)(0){0.5}
\ArrowLine(95,70)(55,70) \Text(75,78)[b]{$b$}
\ArrowLine(55,30)(95,30) \Text(75,25)[t]{$s$}
\Line(10,52)(40,52) \Text(15,55)[b]{$B_s(\bar b s)\to$}
\Line(40,48)(10,48)
\ArrowLine(135,50)(108,63) \Text(115,55)[t]{$c$}
\ArrowLine(95,30)(135,30) \Text(115,25)[t]{$s$}

\DashLine(95,70)(110,85){2}  \Text(102,78)[]{+}
\Photon(95,70)(108,63){2}{2.5} \Line(95,70)(108,63)
\Photon(110,85)(123,79){2}{2.5} \Line(123,79)(110,85)
\DashLine(108,63)(123,79){2} \Text(116,71)[]{+}

\GOval(140,40)(15,7)(0){0.5} \Text(147,40)[l]{$\to D_s^-(\bar cs)$}
\ArrowLine(123,79)(135,73) \Text(128,70)[t]{$u$}
\ArrowLine(135,97)(110,85) \Text(120,95)[b]{$s$}
\GOval(140,85)(15,7)(0){0.5} \Text(147,85)[l]{$\to K^+(\bar s u)$}
\Text(95,0)[b]{$\sim \delta^{d*}_{23}\delta_{12}^u$}
\end{picture}

\vskip 4cm
\begin{itemize}
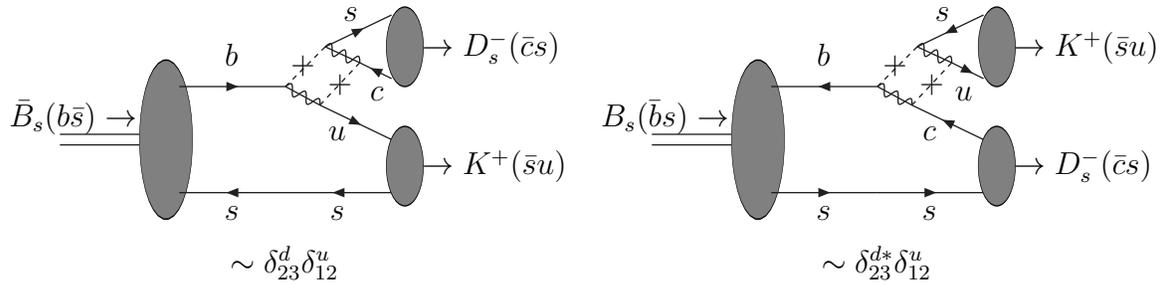
\item[Fig.~2] The Box-graph contribution in SUSY via squark mixing to
the process $(B_s/\bar B_s)\to K^+D_s^-$.
\end{itemize}

\end{document}